# Design and evaluation of a FPGA-ADC prototype for the PET detector based on LYSO Crystals and SiPM arrays

Cong Ma, Xue Dong, Li Yu, Wubin Wang, Xiaokun Zhao, Xing Li, Zhenqing Huang, Guocheng Wu, Lei Lu, Hansheng Chen

**Abstract— the aim of this study is to design and evaluate a simple free running Analog-Digital Converter (ADC) based on the Field Programmable Gate Array (FPGA) device to accomplish the energy and position readout of the silicon photomultiplier (SiPM) array for application as PET scanners. This simple FPGA-ADC based on a carry chain Time-Digital Converter (TDC) implemented on a Kintex-7 FPGA consists of only one off-chip resistor so it has greater advantages in improving system integration and reducing cost than commercial chips. In this paper, a FPGA-ADC based front-end electronics prototype is presented, and both the design principle and implementation considerations are discussed. Experiments were performed using an $8 \times 8$ (crystal size: $4 \times 4 \times 15$ mm³) and a $12 \times 12$ (crystal size: $2.65 \times 2.65 \times 15$ mm³) segmented LYSO crystals coupled with an $8 \times 8$ SiPM (J-series, from ON Semiconductor) array which is under $^{22}$Na point source excitation. Initial test results indicate that the energy resolution of the two detectors after correction is around 13.2% and 13.5 % at 511 keV, and the profiles of the flood histograms show a clear visualization of the discrete scintillator element. All measurements were carried out at room temperature (~25 °C), without additional cooling.**

## I. INTRODUCTION

Nowadays, Positron Emission computed Tomography (PET) based on Silicon Photomultipliers (SiPMs) is widely used as an advanced nuclear medicine imaging technique which is with low operation bias voltage ($V_{bias}$), insensitive to magnetic fields and low transit time spread performance [1-4]. PET instrument exploits the coincidence detection of 511 keV annihilation photon pairs to infer their along lines of response (LOR) between scintillation crystal elements in the system detector ring [5-6]. A PET detector module needs to measure the arrival time, energy and position of the incident 511 keV photon. To identify the tiny scintillator elements for high spatial resolution, a general PET scanner requires tens of thousands of readout channels [7-8]. Power consumption and density become challenging issues when the readout electronics system scales up [9]. An attractive solution is to employ a highly integrated Application Specific Integrated Circuit (ASIC), which could accomplish the measurement at the front end with low power consumption [10-15]. However, ASIC has long research and development (R&D) cycle and high manufacturing cost. Thus, in some applications, light-sharing technique based on a multiplexing network with discrete components is still a common choice to reduce the number of signal processing channels [16-19]. Even so, the number of readout channels is still very large for a general PET scanner. The large number of readout channels in these designs make it necessary to use complex electronics to process signals due to the large number of readout channels. Therefore, there is a strong demand for simple and efficient readout electronics to use in the development of a compact and cost-effective PET system.

Fortunately, researches in recent years succeed in integrating high-resolution time-digital converters into the Field Programmable Gate Array device (FPGA-TDC) which makes the time measurement much simpler [20-23]. Nevertheless, the energy measurement is commonly based on a commercial high-speed analog-digital converter (ADC) chip to digitalize the analogue signals, which certainly leads to high system power consumption and cost [24]. So far the most common alternative solution is time-over-threshold (TOT) method which compares the integral signal with a pre-set threshold to translate the signal amplitude into a TOT time width measured with TDC [25-27]. Reference [26] presents the energy resolution of around 10.4% based on a $4 \times 4$ array of 3 mm × 3 mm × 20 mm LYSOs coupled with a $4 \times 4$ array of 3 mm × 3 mm SiPMs using a bipolar TOT circuit. For some one-to-one readout or hybrid charge division systems, conventional TOT can be used because the pulse shape does not change with respect to position [27]. However, its drawbacks of nonlinearity and small dynamic range limit its scope of application especially in light-shared systems [28]. To optimize its performance, many similar structures are presented, such as time over dynamic threshold (TODT) [29] and fast discharging TOT method [30]. Another common method is called multi-voltage threshold (MVT) for digitizing a PET event by sampling with respect to certain reference voltages. By fitting the mathematical model of SiPM signals to the several sampled pulses, a high energy resolution with low dead time can be obtained [28, 31-32]. Reference [32] presents the average energy resolution of around 11.3% based on 3 mm × 3 mm SiPMs 1:1 coupled with a 20 mm-thickness LYSO crystal using MVT circuit. Nevertheless, MVT circuit consumes lots of comparators or FPGA IOs. Thus, it is still a great challenge to achieve high resolution energy measurement over a large dynamic range without an ADC chip.

In this paper, we designed and optimized a front-end electronics for SiPM signals readout based on a free running ADC implemented on FPGA (FPGA-ADC) first proposed in Reference-[33]. The FPGA-ADC based on a carry chain TDC consists of only one off-chip resistor so it has greater advantages in improving integration than the commercial chips, and electronics tests were performed to assess its dynamic performance, linearity and resolution. Furthermore, experiments were conducted using an 8 × 8 (crystal size: 4 × 4 × 15 mm$^3$) and a 12 × 12 (crystal size: 2.65 × 2.65 × 15 mm$^3$) LYSO crystals coupled with an 8 × 8 SiPM (J-series, from ON Semiconductor) pixel segmented array which was under $^{22}$Na excitation to assess its feasibility in one-to-one coupled and light-shared detectors. The initial evaluation results give us a possibility to develop a FPGA-only front-end digitizer for TOF-PET in future. This paper is organized as follows: Section-*II* introduces the design principle and details of the front-end electronics prototype; in Section-*III*, electronics tests are conducted to evaluate the performance of the readout electronics, and shows the initial experiment results in one-to-one and light-shared detectors, respectively; a brief summary of this work and the discussion about the improvements at next stage are given in Section-*IV*.

## II. MATERIALS AND METHOD

### A. Structure of the multiplexing network

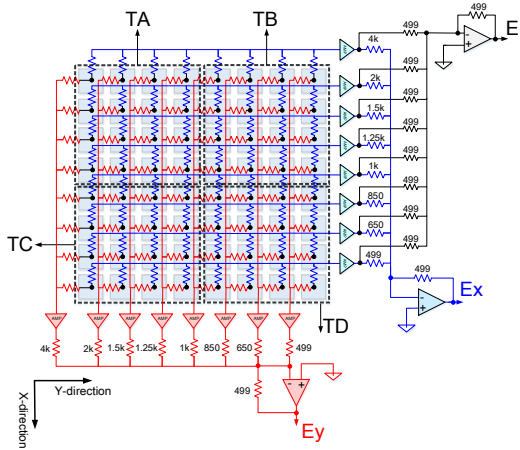

Fig. 1 Structure of the multiplexing network.

The Structure of the multiplexing network is shown in Fig. 1. The 8 × 8 SiPM array outputs 64-channel current analogue signals from anodes. Each channel is split to two pathways by two resistors (1 kΩ) and a symmetric multiplexing circuit ties the 8-X and 8-Y coordinate signals, separately. So, the number of the readout channels is reduced from 64 to 16. By summing the 8-X signals, the charge for each incident 511 keV photon can be collected and then the energy information (E) can be calculated by the following digitalization process. To infer the position of the inspired scintillator pixel, the outputs from all rows (or columns) are amplified with a set of optimized gradient gains ($G_g$) by an analog inverting adder (ADA4857, from ADI Inc.) to acquire the position energy signals ($E_x$ and $E_y$). Theoretically, the flood histogram can be generated using the below decoding logic [34]:

$$X = \frac{E_x}{E} \quad Y = \frac{E_y}{E} \tag{1}$$

To identify each row or column of crystal array, we optimized $G_g$ configurations for our prototype. In general, the measured energy resolution (*Res*) is mainly affected by the following factors: the crystal and SiPM intrinsic energy jitter ($\sigma_0$), the noise jitter from the electronics components ($\sigma_{noise}$), the jitter from the A/D conversion ($\sigma_{ADC}$) and the digital integration error ($\sigma_{int}$) [35]:

$$Res = \frac{\sqrt{\sigma_0^2 + \sigma_{noise}^2 + \sigma_{ADC}^2 + \sigma_{int}^2}}{A};\tag{2}$$

where, *A* is the mean of the measured energy.

For a traditional readout system, the effect introduced by the electronics ($\sigma_{noise}$, $\sigma_{ADC}$ and $\sigma_{int}$) can be ignored compared with $\sigma_0$, and thus the signals amplified with different gains from $E_x$ or $E_y$ have basically the same resolution performance for an incident 511 keV photon. Therefore, in order to distinguish each full width at half maxima (*FWHM*) in the energy spectrum of the $E_x$ or $E_y$, $G_g$ should be set as -1,-1/2,-1/3,-1/4,-1/5,-1/6,-1/7 and -1/8. However, as mentioned in the following sections, the performance of the FPGA-ADC employed in our design cannot compare favorably with the high-resolution commercial chips. This leads to larger $\sigma_{ADC}$ and for small signals, the energy resolution would be deteriorated. Therefore, it is necessary to increase the gain gradients at the low gain end for getting a clearer position identification. By simulations and tests, the $G_g$ is finally optimized to -1,-0.77,-0.59,-0.5,-0.4,-0.33,-0.25,-0.125. The uneven gain difference can be further corrected in the digital signal processor.

Meanwhile, for the J-series SiPM, a fast output is specifically designed to accomplish high-resolution time measurement [36]. In our design, to reduce the baseline fluctuation caused by the accumulation of dark counts and discrete capacitance originated from multiple SiPMs, quarter of the SiPMs are separately added to obtain four timing signals (TA,TB,TC and TD), and then discriminated by four leading edge discriminators (LEDs). Then the four outputs are summed to one time measurement channel by a OR gate (74LVX32M, from ON Semiconductor) to get the first coming leading edge. Thus, we can get the arrival time of the coming gamma ray.

### B. Digital signal processor

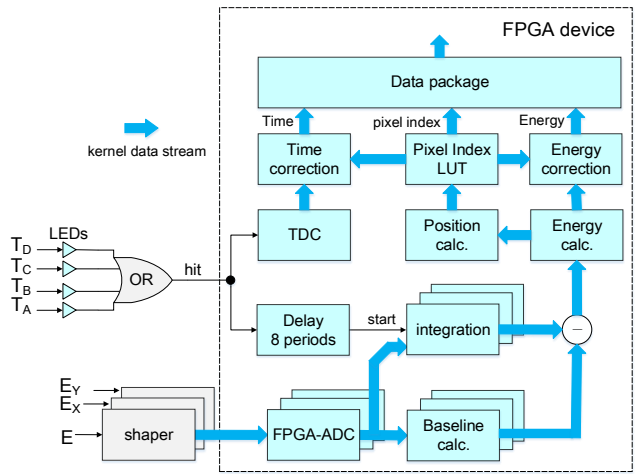

Fig. 2 Structure of the digital signal processor.

The digital signal processor as shown in Fig. 2 has been implemented on a Kintex-7 FPGA (XC7K325T-2FFG900, from Xilinx Inc.). For the time measurement, a carry chain FPGA-TDC with an effective bin size of around 20 ps is employed. Previous research indicated that the resolution of this type TDC could be better than 60 ps FWHM [37], which is good enough for the TOF-PET application. For the energy measurement, the analogue signals from E, $E_x$ and $E_y$ are shaped to a Quasi Gaussian pulse by a CR-RC$^3$ filter and then fed into the FPGA-ADCs. Then, the digitalized data stream is integrated during several sampling periods (~600 ns) to obtain the energy information. According to Equation-1, the position flood histogram can be calculated and the index of the inspired scintillator pixel can be queried by a pre-generated look-up-table (LUT). Once the index of the inspired pixel is determined, the energy and time measurement results can be corrected by the parameter LUTs stored in FPGA. The details about the on-line pixel index query and correction procedure have been reported in Reference-[38]. In Section III-D, we also illustrate the basic principle combined with the actual test data. Moreover, the event-by-event baseline correction is done using the mean value of 8 data points before the onset of the signal. Finally, the energy, position and time information are packed and transmitted to the back-end data acquisition (DAQ) and host computer.

The focus of this work is to discuss the full Hardware Description Language (HDL) based FPGA-ADC, whose implementation details are shown in the next Sub-section.

*C. ADC implemented on FPGA*

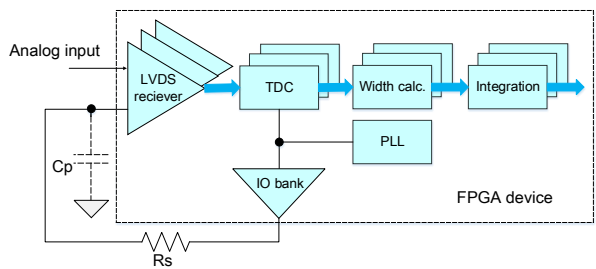

Fig. 3  ADC implemented in FPGA.

The structure of the FPGA-ADC presented in this work is shown in Fig. 3.  A clock synthetized by the internal phase-locked loop (PLL) is filtered to a quasi-triangular sampling ramp by a simple low pass filter formed by a serial resistor ($R_s$) with the parasitic capacitance ($C_p$) at the input of the user I/O ports. The analogue input signal is discriminated by the sampling ramps using the differential FPGA I/Os working in LVDS receiver mode [33] and the width of the output digital pulses can be used to characterize the analog signal voltage amplitude. A carry chain TDC as mentioned in Section II-B is employed to achieve the pulse width measurement. The high level of output clocks is 3.3 V.

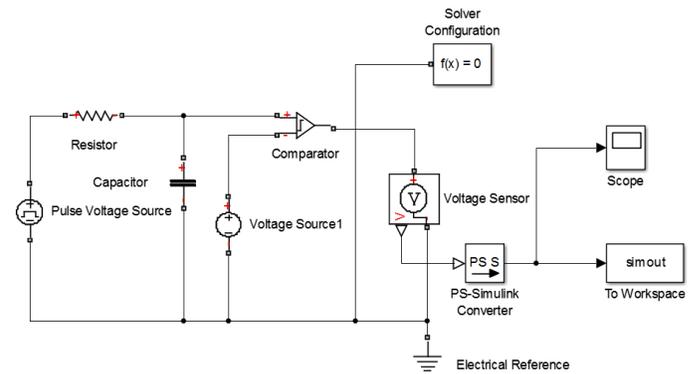

Fig. 4 (a). The Matlab Simulink simulation platform.
(kernel parameters: $F_s$ is 25 MHz; $F_{in}$ is 1 MHz; $R_s$ is 90 ohms; $C_p$ is  180 pF; simulation time step is 40 ps. All the electrical components are based on ideal model, without electrical noise)

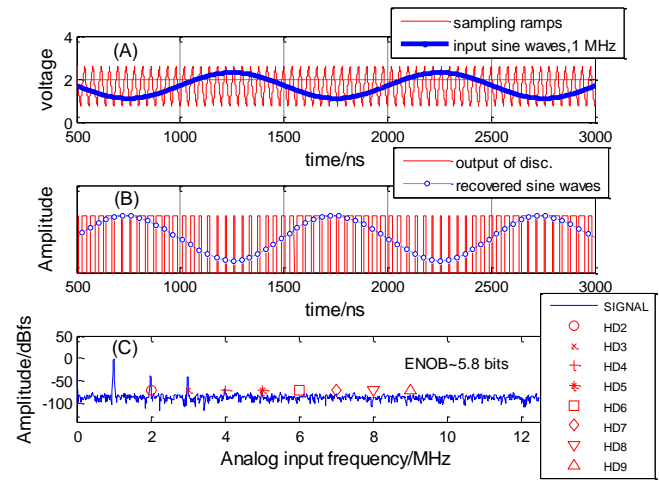

Fig. 4 (b). Transient waveforms and frequency spectrums simulation results.
(A: the transient waveforms of the sampling ramps and input sine waves;
B: the transient waveforms of the digital pulses and the recovered sine waves;
C: the frequency spectrums of the recovered sine waves.)

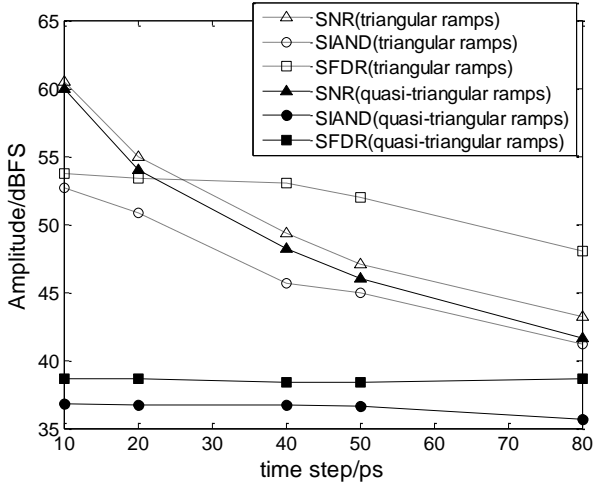

Fig. 4 (c). Dynamic performance simulation comparison between quasi-triangular and triangular sampling ramps. (kernel parameters: $F_s$ is 25 MHz; $F_{in}$ is 1 MHz; $R_s$ is 90 ohms; $C_p$ is 180 pF. All the electrical components are based on ideal model, without electrical noise)

One of the important factors affecting performance is that the sampling frequency ($F_s$) and $R_s$ should be properly chosen to balance the measurement resolution and measureable range. For high sampling frequency or large $R_s$, the long time constant of the low pass filter ($R_s \times C_p$) would seriously attenuate the sampling ramps and deteriorate the measurable range. On the other hand, for low sampling frequency or small $R_s$, the sampling ramps would be much more similar to exponential ramps rather than the triangular ramps, which would lead to large conversion nonlinearity. Besides, for the narrow nuclear signals, low sampling frequency would increase the digital integration error ($\sigma_{int}$). We optimized it by simulations on Matlab Simulink platform, as shown in Fig. 4. $F_s$ and $R_s$ are configured as 25 MHz and 90 ohms, and the measurable amplitude range can reach around 1.7 V (0.8 V ~ 2.5 V). The simulation sweep time step is set 40 ps to imitate the behavior of TDC. The FPGA-ADC digitalizes the input sine waves with frequency ($F_{in}$) of 1 MHz and by Fourier transform we can analysis its dynamic performance. Fig. 4(b) shows that the effective number of bits (ENOB) is around 5.8 bits, and the harmonics at the multiples of the input frequency have great influence. In this circuit, one of most important reasons introducing the harmonics is that the sampling ramps are actually not real triangular waves, which causes serious measurement nonlinearity. A simple mathematical model can give a brief explanation. For a nonlinear system, the output ($V_{out}$) can be expressed as Equation-3. Obviously, several harmonics at the multiples of the input frequency ($\omega, 2\omega ...$) will occur on the output. [39]

$$V_{out} = a_0 + a_1 V_{in} + a_2 V_{in}^2 + a_3 V_{in}^3 + \cdots$$
$$V_{in} = \cos(\omega t)$$
$$\cos(\omega t)^2 = \frac{1+\cos(2\omega t)}{2} \qquad (3)$$

To verify this, we compare the simulated dynamic performance using real triangular and the quasi-triangular sampling ramps, as shown in Fig. 4(c). The simulation time step is from 10 ps to 80 ps and the SNR, Signal to Noise and Distortion Ratio (SINAD) and Spurious-Free Dynamic Range (SFDR) are calculated [40]. Results show that, compared with the ideal sampling circuit using standard triangular ramps, the deteriorations of SFDR and SINAD are much more obvious than SNR in the actual circuit. In other words, for the same time step configuration, the harmonics introduced by the sampling ramps are the major factor of deteriorating the dynamic performance.

Besides, Fig. 4(c) also indicates that the quantization noise introduced by the TDC directly affects the ADC's SNR performance. For the carry chain TDC, the fine delay tap should be designed as small as possible and an online nonlinearity correction and temperature compensation mechanism should be implemented. In our design, the TDC is normalized to 9 bits after correction with an equivalent bin size of around 7.8 ps. Considering the maximum width of the measured pulses is 40 ns, the range of the ADC is larger than 12 bits.

Moreover, in our application, the dispersion of the sampling ramps among channels should be considered. Since the large uncertainty of the parasitic capacitance at the I/O ports, the sampling ramps for the three position energy signals from the same array should be shared to guarantee the performance uniformity. In addition, this distribution pattern requires us to control the impedance match and length of the transmission line for better signal integrity.

Above all, the main design parameters, performance simulation results and resource usage of the FPGA-ADC are listed in Table 1. To overcome the drawbacks of low sampling frequency and dynamic performance of the FPGA-ADC, a low pass shaper is employed to shape the SiPM signal to a wider Quasi Gaussian pulse and the energy integration time is around 600 ns.

Table 1 the main design parameters, performance simulation results and FPGA resource usage per channel of the FPGA-ADC.

| Item | performance |
|---|---|
| simulated measurable amplitude range | 0.8~2.5 V |
| sampling rate | 25 Msps |
| resolution | >12 bits |
| ENOB simulation results | ~5.8 bits@1MHz |
| IO usage | 2/ch |
| LUT usage | 5574/ch |
| FF usage | 4156/ch |

*D. Detectors configurations*

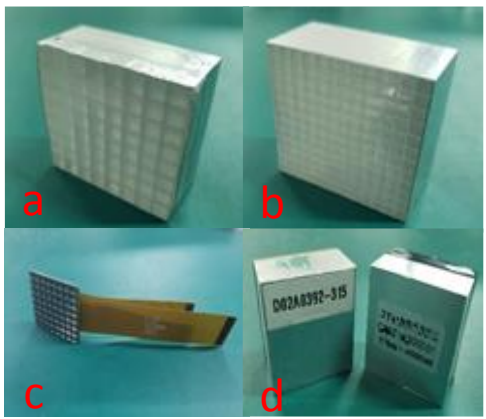

Fig. 5 The tested LYSO crystals and SiPM array.
(a: an 8×8 LYSO crystal, size: 33.8×33.8×15 mm³;
b: a 12×12 LYSO crystal, size: 33.8×33.8×15 mm³;
c: 8×8 SiPM array with the carry board and flat cables;
d: two coupled detectors.)

The SiPM pixel to be evaluated has an active area of 3.93 × 3.93 mm², and the 8 × 8 array has a total area of 33.8 × 33.8 mm² (pitch size: 4.2 mm). To ensure a stable operating temperature of the SiPM array, a carry board consisting of the SiPM array and the multiplexing network without any other active devices is designed and it is connected with the front-end board via flat cables. The operating bias voltage for all pixels on one module is 29.5 V.

In the experiment, two pixel segmented LYSO crystals are coupled to the SiPM array. One is an 8 × 8 scintillator array with a pixel size of 4.0 mm (pitch size: 4.2 mm) and thickness of 15 mm to test the one-to-one coupled detector. Other is a 12 × 12 scintillator array with a pixel size of 2.65 mm (pitch size: 2.8 mm) and thickness of 15 mm to test the light shared detector, and the thickness of the light guide glass between crystal and SiPMs is 1.0 ± 0.03 mm. All the LYSO pixels has been polished and later covered by reflective material (Teflon). The LYSO, light guide and the SiPM are pasted with optical grease (BC630).

The experiments are under a $^{22}$Na point source (~60 μC$_i$) excitation, and a $^{137}$Cs point source (~12 μC$_i$) is used to conduct the saturation correction of the SiPMs.

## III. TEST PLATFORM AND RESULTS

### A. Evaluation platform

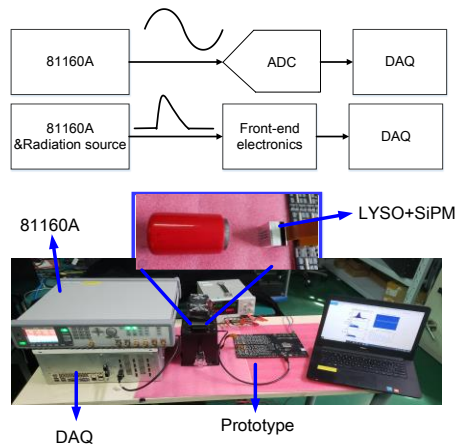

Fig. 6 System under test.

The evaluation platform in the laboratory is shown in Fig. 6. Firstly, an arbitrary signal generator (81160A, from Keysight Inc.) outputs user defined signals to independently assess the FPGA-ADC's dynamic performance (shown in Section *III-B*), and then the measurement linearity and resolution of the front-end electronics (shown in Section *III-C*). Secondly, two LYSO crystals mentioned above coupled with the SiPM array are tested under $^{22}$Na excitation (around 10 cm distance) to acquire the energy spectrum and flood histogram (shown in Section *III-D* and *III-E*).

### B. Dynamic performance of the FPGA-ADC

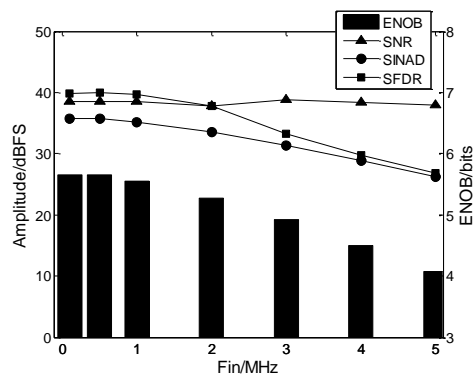

Fig. 7 Dynamic performance of the FPGA-ADC.

The arbitrary waveform generator outputs continuous sine waves with a nearly full voltage swing to the FPGA-ADC, and the dynamic performance is shown in Fig. 7. The results indicate that the ENOB is around 5.5 bits at 1 MHz and 4 bits at 5 MHz. The test results of SNR is worse than the simulation probably because of the effects of circuit noise and the TDC's nonlinearity, etc.

### C. measurement conversion curve and resolution

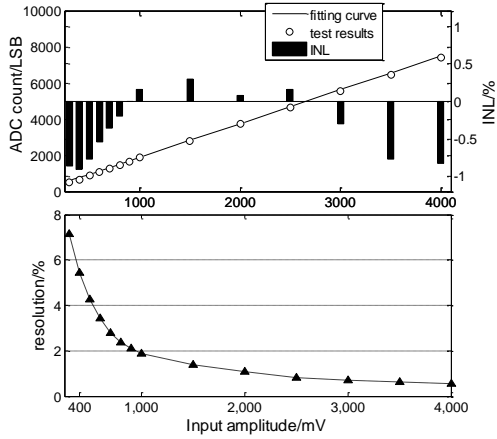

Fig. 8 integration conversation curve and resolution test results.

Considering it is necessary to measure the position energy signals over a large dynamic range in light shared application, we tested the linearity and resolution of the front-end electronics. The arbitrary waveform generator outputs SiPM-like signals (rise time ~ 50 ns, fall time ~ 200 ns) to the circuit. The signals are shaped and then sampled by the FPGA-ADC.

Similar to the simulation, the ADC's measurable voltage range is from around 0.8 V to 2.5 V. In order to ensure the measurement stability of the baseline, the shaped signals are designed with a DC bias of 0.9 V. Fig. 8 shows the measurement conversation curve and the measured resolution at different circuit input voltage amplitude. Test results indicate that the measurement integral nonlinearity (INL) is within ±1.0% without any correction and the resolution is better than 6.0% in the input range from 400 mV to 4 V. At the input port of the FPGA-ADC, this range is from 110 mV to 1 V with a 0.9 V DC bias.

### D. One-to-one Coupled LYSO/SiPM detector

The 8×8 LYSO crystal mentioned above was coupled to the SiPM array to test the energy spectrum and flood histogram in the one-to-one application. Since the dispersion of the gain of each pixel and the uneven gain gradient of the multiplexer, the following correction scheme as shown in Fig. 9 is conducted. Firstly the raw flood histogram is discretized to 512 image pixels, and then filtered and clustered to 8 × 8 barycenter positions of crystal pixels. By Voronoi division, the crystal pixel index and segmentation area can be obtained. The boundaries of the divisions in X and Y direction are then respectively stored as LUTs (size of each: 512 × 8 × 9 bits ) in FPGA, which can be used to query the inspired crystal pixel index according to the raw position energy value. Reference-[38] shows more details about it.

The next is to get the energy correction parameters of each pixel. In order to calibrate the nonlinearity of the SiPMs, a $^{137}$Cs point source was also tested. We used the below model to fit the energy peak at 511 keV and 662 keV of each pixels and get the correction parameters ($b$, $k$) [41]. In addition, for $E_x$ and $E_y$, the row and column electrical gradient gains were also compensated to make the distribution of flood histogram even.

$$p = -N \cdot \ln\left(1 - \frac{b \cdot k}{N}\right) \qquad (4)$$

where $p$ is the number of photons that would have been detected without saturation; $k$ is the measured energy integration results (ADC count); $b \cdot k$ is the number of triggered cells with saturation; $N$ is the limited number of cells in an SiPM pixel.

The energy spectrums before and after correction are shown in Fig. 10. Test results indicate that the energy resolution after correction is around 13.2 %.

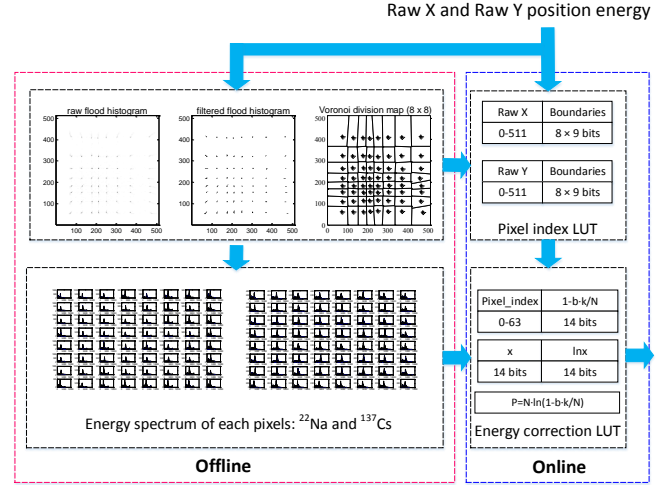

Fig. 9 procedure of energy correction.

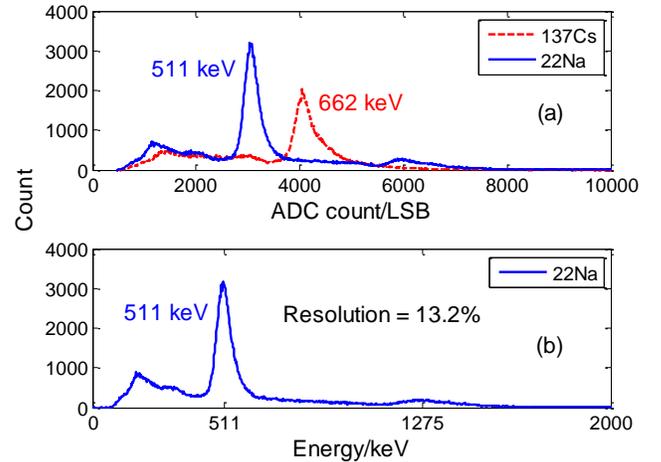

Fig. 10 uncorrected (a) and corrected (b) energy spectrum of the one-to-one detector.

The flood histogram after energy correction is shown in Fig. 11, with a 425 ~ 650 keV energy window. The corrected energy spectrum has pretty good resolution performance, and the eight peaks of $E_x$ or $E_y$ can be distinguished clearly. To quantize the image quality, the distance-to-width ratio (DWR) is calculated as the average ratio between the distance of the adjacent peaks on the flood histogram and the average FWHM value of the peaks, as described in Equation-2 [42].

$$DWR = \frac{1}{N_{adj}} [\sum_{i,j \in (adj\ pair)}^{N_{adj}} \left(\frac{|x_i - x_j|}{\frac{w_{x,i} + w_{x,j}}{2}}\right) + \left(\frac{|y_i - y_j|}{\frac{w_{y,i} + w_{y,j}}{2}}\right), (2)$$

where $x_i$, $x_j$ and $y_i$, $y_j$ correspond to the position of the $i^{th}$ and $j^{th}$ adjacent crystal pairs along the horizontal axis (x-axis) and the vertical axis (y-axis) on the flood histogram, respectively. $w_{x,i}$, $w_{x,j}$ and $w_{y,i}$, $w_{y,j}$ are the FWHM values of 1-D projection profiles along the x- and y-axis of the $i^{th}$ and $j^{th}$ crystal, respectively. $N_{adj}$ is the total number of adjacent crystal pairs.

The DWR for the one-to-one flood histogram is around 11.0.

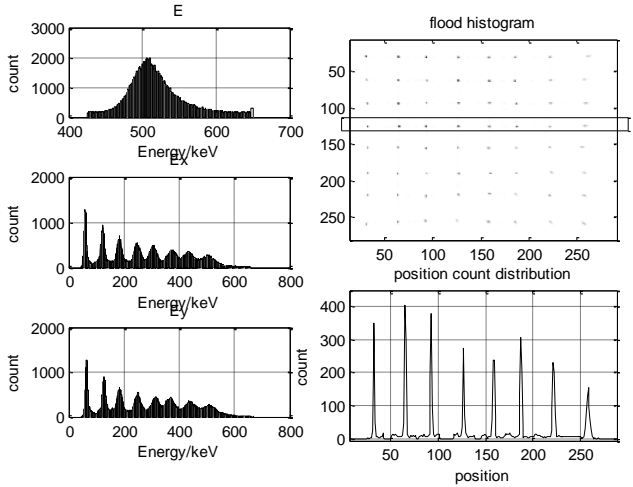

Fig. 11 corrected energy spectrum of E, $E_x$ and $E_y$, as well as the flood histogram of the one-to-one detector.

### E. Light-shared LYSO/SiPM detector

Another 12 × 12 LYSO crystal was then coupled with the same SiPM array which is under $^{22}$Na excitation to test this prototype applied in light-share detector. The uncorrected and corrected energy spectrums of the module are shown in Fig. 12, and the resolution is around 13.5 %. The corrected histogram are shown in Fig. 13, with a 425 keV ~ 650 keV energy window. The DWR of this detector is around 7.8.

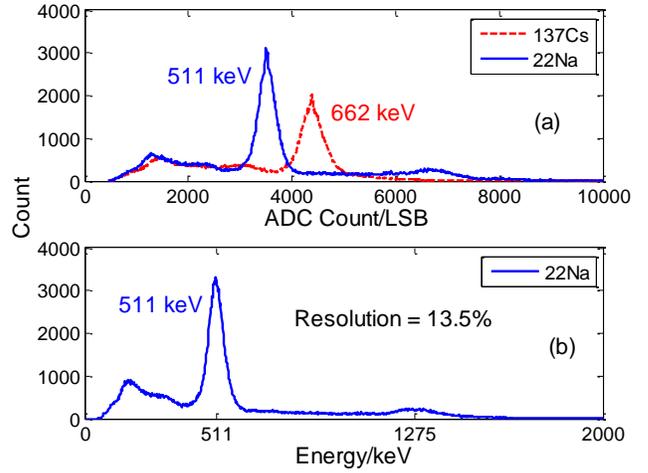

Fig. 12  uncorrected (a) and corrected (b) energy spectrum of the light-share detector.

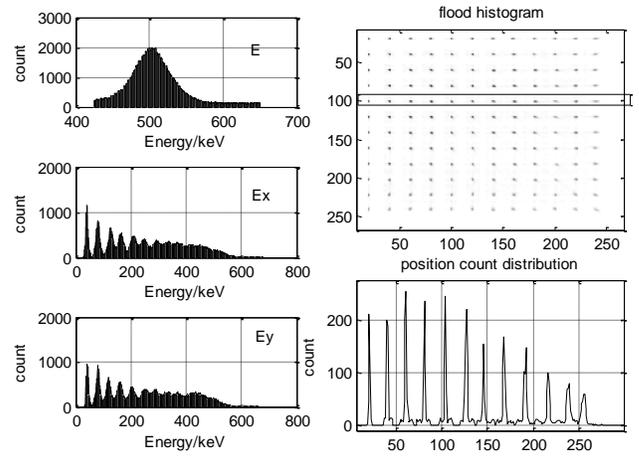

Fig. 13 corrected energy spectrum of E, $E_x$ and $E_y$, as well as the flood histogram of the light-share detector.

### F. Time resolution performance

Using the timing signal, we measured the coincidence time spectrum between two detectors with same configuration. The timing threshold was experimentally optimized and the time-walk correction of the leading edge timing pick-off was performed using a linear correlation between the measured energy and time values. The time propagation delay difference of all pixels were calibrated.

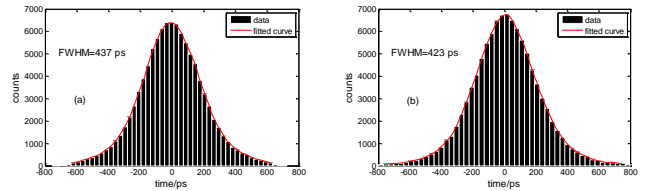

Fig. 14 coincidence timing difference of two 8×8 crystal detector (a) and 12×12 crystal detector(b).

Table 2 electronics performance and main parameters of several reported FPGA-based measurement methods (per channel).

| Method<br>Item | TOT [25] | TODT [29] | Fast discharging TOT [30] | MVT [28] | FPGA-ADC in this paper |
|---|---|---|---|---|---|
| dynamic range | 0.1 V ~ 2 V | 0.1 V ~ 1 V | 700 mV ~ 1300 mV | — | 0.11V ~ 1V |
| Resolution (FWHM) | <15% | — | — | ~6 % | < 6 % |
| Max Dead time | ~1 us | ~400 ns | ~300 ns | <100 ns | ~600 ns |
| Main auxiliary components | One DAC chip | One high-speed DAC | One OPA | One DAC chip | One resistor |
| FPGA I/O usage | 1 | 2 | 3 | 8 | 2 |

One typical time resolution test result between two $8 \times 8$ LYSO crystal coupled detectors is shown in Fig. 14(a), where the FWHM is fitted as 437 ps. Since the timing performance of the coincidence detector is known, the time resolution for two identical test detectors can be estimated as 309 ps. Similarly, Fig. 14(b) shows that the FWHM is around 423 ps between two $12 \times 12$ LYSO crystal coupled detectors, with the resolution of 299 ps for two identical detectors.

## IV. DISCUSSION AND CONCLUSION

### A. Discussion and future work

#### 1) Electronics performance comparisons

The electronics performance and parameters of several reported FPGA-based measurement methods are listed in Table 2. The FPGA-ADC presented in this paper has some advantages of dynamic range and integration. The further optimization mainly towards to increasing the dynamic and resolution performance.

#### 2) Nonlinear correction of the FPGA-ADC

Although the detector experiment results mentioned above show that the FPGA-ADC could be an alternative scheme for some energy measurement applications, the drawbacks of its low sampling rate and unsatisfactory performance obviously limits its applications. For example, as mentioned in Section *III-C*, the energy measurement INL of the front-end electronics is within ±1.0% and the resolution is better than 6.0% in the input range from 400 mV to 4 V. This is basically enough for a light-shared detector with a $12 \times 12$ crystal coupled with $8 \times 8$ SiPM array evaluated in this paper. However, for a detector with more tiny crystal element size, the energy measurement requirements of dynamic range and small signals resolution are much tighter, and the FPGA-ADC may be unusable. Therefore, the next kernel work is how to increase the ADC's ENOB performance.

As mentioned in Section *II-C*, The ADC's dynamic performance is mainly limited by the measurement nonlinearity introduced by the sampling ramps. The simulation shows that if the nonlinearity is completely eliminated, the ENOB could be larger than 7 bits. This will broaden the application range of this type ADC. Following this way, one possible solution is to calibrate the function of the sampling ramps and make a LUT stored in FPGA to correct the nonlinearity. This work is currently under study and would be reported in future.

#### 3) Time interleaved FPGA-ADC

As foreside, the designed FPGA-ADC's sampling frequency is only 25 MHz, and we have to shape the narrow analogue signals from SiPMs. Long energy integration time would lead the circuit cannot recovery the pile-up signals and a long dead time, which limits its application in high rate cases and deteriorates the noise effective count rate (NECR) performance for PET scanner [43]. One possible solution is to employ two channels with sampling clocks of 180° phase shift to achieve time interleaved modulation. By this method, the sampling rate will be doubled but a correction mechanism is needed to correct the gain, phase and offset difference between the channels [44].

### B. Conclusion

Preliminary feasibility of the FPGA-ADC applied in the readout electronics for SiPM signals has been discussed in this paper. This simple FPGA-ADC based on a carry chain Time-Digital Converter (TDC) implemented on a Kintex-7 FPGA consists of only one off-chip resistor so it has great advantages in improving system integration and reducing cost. We designed and manufactured a front-end electronics prototype to verify if the simple FPGA-ADC could be used to digitalize the SiPM signals. Experiments were performed using an $8 \times 8$ (crystal size: $4 \times 4 \times 15$ mm$^3$ ) and a $12 \times 12$ (crystal size: $2.65 \times 2.65 \times 15$ mm$^3$ ) LYSO crystals coupled with an $8 \times 8$ SiPM (J-series, from ON Semiconductor) array which is under $^{22}$Na excitation. The initial test results indicate that the typical energy resolution of the two detectors after correction is around 13.2% and 13.5 % at 511 keV, and the profiles of the flood histogram shows a clear visualization of the discrete scintillator element. Besides, the application limitations, drawbacks and the direction of future improvements are also mentioned.